\documentclass[journal]{vgtc}                     
\usepackage{cite}
\usepackage{amsmath}
\usepackage{amssymb}
\usepackage{booktabs} 
\usepackage{subcaption}

\usepackage[percent]{overpic}

\usepackage[many]{tcolorbox}

\definecolor{sub}{HTML}{e3e3e3}

\tcbset{
    sharp corners,
    colback = white,
    before skip = 0.2cm,
    after skip = 0.5cm
}
\newtcolorbox{grammarBox}{
    colback = sub,
    boxrule = 0pt
}

\onlineid{1624}

\vgtccategory{Research}

\vgtcpapertype{System}

\newcommand{\sysname}{LatentGandr\xspace}

\title{\sysname: Visual Exploration of Generative AI \\Latent Space via Local Embeddings}

\author{%
  \authororcid{Mingwei S.G. Li}{0000-0002-0457-8091}
  \authororcid{Susie S.Y. Li}{0009-0000-6375-182X}
  \authororcid{Daisuke Sakurai}{0000-0002-0464-8619
}
  \authororcid{Bei Wang}{0000-0002-9240-0700}
  \authororcid{Remco Chang}{0000-0002-6484-6430}
}

\authorfooter{
  \item
  	Mingwei S.G. Li, Susie S.Y. Li and Remco Chang are with Tufts University.
    E-mail: \{mingwei.li, suyang.li, remco.chang\}@tufts.edu
  \item
  	Daisuke Sakurai is with Fujitsu.
  	E-mail: sakurai.daisuke@fujitsu.com
   \item
  	Bei Wang is with the University of Utah.
  	E-mail: beiwang@sci.utah.edu
}

\abstract{
Generative AI has demonstrated significant potential in creative design, enabling the rapid generation of visual content and imaginative concepts. 
Although deep AI models achieve effective featurization in the latent space, navigating the space remains a challenge. 
Current techniques, such as GANSlider and SliderSpace, use multiple sliders to generate high-dimensional vectors in generative AI’s latent space. 
Despite applying (global) PCA to reduce the number of sliders, these approaches struggle with scalability and usability as the number of control dimensions increases.
In this paper, we introduce \sysname, a visual analytics technique that facilitates latent space exploration by extracting locally linear dimensions from embeddings in high-dimensional latent spaces. 
By analyzing the topology and local curvature of the embeddings, \sysname automatically identifies local neighborhoods and computes their principal components using localized PCA. 
These local principal components are visualized as interactive image grids, allowing users to efficiently explore and control the generative process, providing an intuitive means to refine the generation of novel content and concepts.
To evaluate the effectiveness of \sysname, we conducted a study comparing it to GANSlider, the current state-of-the-art visualization interface for generative AI models. 
The results offer insights into how localized exploration techniques can enhance user interaction with these models.

}


\teaser{
    \centering
    \includegraphics[width=1\linewidth]{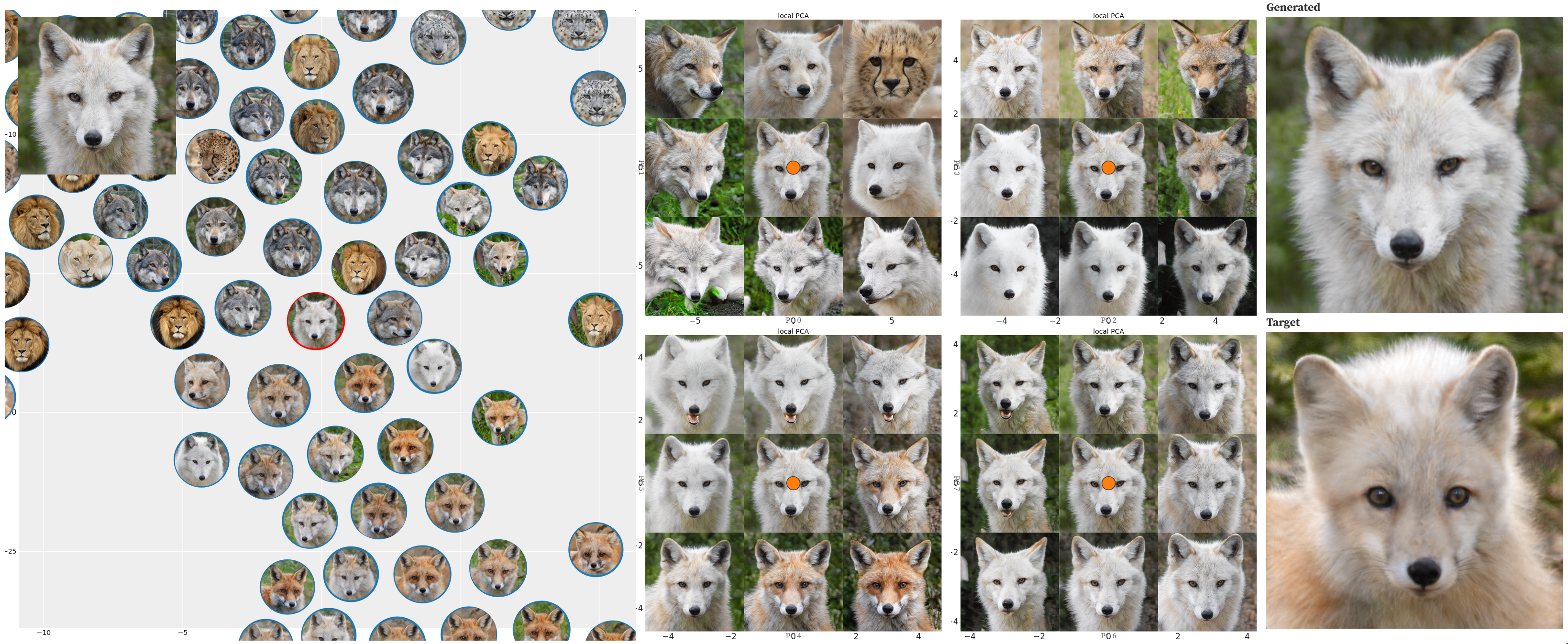}
    \caption{
    \sysname interface for localized latent space exploration. Left: graph layout of latent space with local neighborhood; center: image grids showing variations along local principal components; right: generated image vs. target.
    }
    \label{fig:teaser}
}

\graphicspath{{figs/}{figures/}{pictures/}{images/}{./}} 

\usepackage[english]{babel}
\usepackage[pagebackref,bookmarks]{hyperref}
\addto\extrasenglish{

}

\usepackage{booktabs}                  
\usepackage{lipsum}                    
\usepackage{mwe}                       
\usepackage{listings}

\usepackage{mathptmx}                  
\usepackage{xspace}

\usepackage{fancybox}

\usepackage{tabularx}

\usepackage{tikz}

\usepackage{amsfonts}
\usepackage{dsfont}
\newcommand{\R}{\mathds{R}}

\usepackage[normalem]{ulem}
\usepackage{xcolor}
\definecolor{mred}{rgb}{.80,.12,.30}
\definecolor{grey}{rgb}{0.5,0.5,0.5}
\definecolor{Purple}{rgb}{.75,0,.85}
\definecolor{BlueGreen}{rgb}{.05,.59,.73}
\definecolor{amber}{rgb}{1.0, 0.75, 0.0}
\definecolor{steelblue}{rgb}{0.27, 0.51, 0.71}
\definecolor{violet}{rgb}{0.56, 0.0, 1.0}
\definecolor{myorange}{rgb}{0.94, 0.36, 0.13}

\newif\ifnotes
\notestrue

\let\origcite\cite

\renewcommand{\cite}[1]{\ifnotes\mbox{\origcite{#1}}\else \origcite{#1}\fi}

\newcommand{\commentinline}[2][_]{\textcolor{red}%
{%
\ifx#1_[
\else[#1: \fi
#2]}
}

\begin{document}


\maketitle

\section{Introduction}

Generative AI models have emerged as powerful tools in various domains, including 
creative content generation. 
These models, such as VQ-VAE~\cite{van2017neural}, StyleGAN2\cite{karras2020analyzing} and Stable Diffusion~\cite{rombach2022high}, can synthesize realistic images and generate 
designs.
Their ability to learn intricate patterns from data has revolutionized fields ranging from computer vision to computational art. 
Modern deep learning models tend to map the high-dimensional input pattern in an intermediate space known as the latent space, whose coordinates efficiently encode the desired features of the output in a lower dimensional space.
Visualization methods have been proposed to facilitate exploration and manipulation of this latent space \cite{dang2022ganslider}. 
However, 
its high dimensionality continues to pose challenges for
intuitive visualization, exploration, and control.

It follows that the latent space of a generative model represents a compressed encoding of data, where each point corresponds to a unique output -- for example, a synthesized face or artwork.
In models like StyleGAN2, this latent space can span 512 dimensions or more, making it nearly impossible for users to manually interpret or navigate the space effectively. To fully utilize the potential of these models, users need visualization interfaces that enable intuitive exploration without demanding expertise in high-dimensional mathematics.

At first glance, this challenge appears to be a conventional high-dimensional visualization problem, where the goal is to apply dimensionality reduction (DR) techniques to project $D$-dimensional data points down to two dimensions:
\begin{equation}
    \text{DR}: \R^D \to \R^2
\label{eq:DR}
\end{equation}

However, in the case of visualizing the latent spaces of generative AI models, the objective is not merely to see high-dimensional data points in 2D. Instead, users must be able to select points in 2D that correspond to meaningful high-dimensional vectors, which can then be passed to a decoder to generate outputs. Thus, the visualization task effectively requires the inverse of the traditional DR problem:
\begin{equation}
    \text{GenAI-VIS}: \R^2 \to \R^D
\label{eq:GenAI-VIS}
\end{equation}

A na\"ive approach to solving this problem is to provide users with $D$ sliders, where each slider controls one dimension of the latent space. This approach can be expressed as:
\begin{equation}
    \text{Na\"ive-Sliders}: \underbrace{\R\times\R\times\R\times ... \times\R}_D \to \R^D
\end{equation}

While this approach ensures users with control of the full $D$-dimensional latent space, it quickly becomes impractical as $D$ increases into the hundreds or thousands, making manual adjustment infeasible. Moreover, interactions between dimensions further complicate navigation, imposing significant cognitive burdens on the users.

More advanced approaches, such as GANSlider\cite{dang2022ganslider}, address this challenge by applying principal component analysis (PCA) to reduce the number of control sliders. In this method, a reduced PCA space (referred to as ``Global PCA'' in this paper) is learned from either the training data of the generative model or a carefully sampled set of points in the latent space. Each principal component (Global-PC) is then assigned a slider, potentially significantly reducing the number of adjustable parameters compared to $D$. This approach can be expressed as:
\begin{equation}
    \text{GANSlider}: \underbrace{\R\times\R\times ... \times\R}_{\text{number of Global PCs}} \to \R^D
\end{equation}

Depending on the topology of the manifold, Global-PCA-based approaches can suffer from two key limitations.
First, there is no guarantee that a point in the Global-PCA space lies close to the actual manifold.
When the manifold is derived from training data, this can lead to a higher degree of \textit{hallucination}, where generated outputs become less reliable. 
Second, Global-PCs may struggle to accurately capture the local curvature and structure of the embedding, potentially requiring a larger number of PCs (and consequently, more sliders) to adequately represent the variations in the data.

In this paper, we introduce \sysname, a visual analytics technique designed to improve exploration and manipulation of high-dimensional latent spaces by utilizing locally linear dimensions derived from the topology and local curvature of embeddings.
\sysname identifies local neighborhoods within the latent space and computes principal components specific to these regions (Local-PCA).
By extracting locally linear dimensions (Local PCs) instead of relying on global dimensionality reduction, \sysname preserves the structure of the embedding, providing users with more intuitive control over the generative process.
Following a similar notation as above, our approach can be expressed as:

\begin{equation}
    \text{\sysname}: \{f_i: \underbrace{\R\times\R\times\R ... \times\R}_{\substack{\text{number of Local PCs}\\ \text{in region}~i}} \to \R^D \text{ for } i=1, 2, 3 \dots C\; \}
\end{equation}

Here, $C$ represents the number of local neighborhoods, where each local neighborhood, $i$, may have a different number of Local PCs. 
The size of each local neighborhood serves as an automatically tunable hyperparameter in \sysname, ensuring that the number of Local PCs within a given local neighborhood (and therefore the number of sliders a user interacts with) remains below a pre-specified threshold.

Beyond the algorithmic component, we propose a visual interface for \sysname that consists of two key components:
(1) an overview that enables the user to select a local neighborhood and
(2) a detailed view that enables a user's selection of a point in the Local-PCA space.
Although the specific visualization design may vary, Figure~\ref{fig:teaser} presents an example where the overview is rendered as a 
graph layout of the local neighborhoods, while the Local PCs upon selection are represented as interactive 
image grids:


\begin{equation}
    \text{\sysname}: \{f_i: \underbrace{\R^2\times\R^2 ... \times\R^2}_{\substack{\frac{1}{2}\text{number of Local PCs}\\ \text{in region}~i}} \to \R^D \text{ for } i=1, 2, 3 \dots C\; \}
\end{equation}

To evaluate \sysname, we conducted four experiments: two to assess its algorithmic performance and two to compare \sysname with GANSlider, the state-of-the-art interface for latent space exploration of generative AI models. Specifically, our first experiment evaluates the fit of the Local PCs extracted by \sysname and their ability to model a local neighborhood within the embedding. The second experiment compares our Local-PCA approach with Global-PCA and UMAP, focusing on the quality of the identified local neighborhoods.

For the comparison with GANSlider, we first assessed the reliability of user-selected points in the Global-PCA space versus the Local-PCA space, measuring their distances to the manifold as a proxy for the likelihood of AI hallucination. We then conducted an in-person experiment where participants performed an image recreation task using both GANSlider and \sysname. 
The results of our experiments suggest that \sysname performs strongly in our algorithmic evaluations, effectively selecting local neighborhoods and extracting local PCs. In the comparison with GANSlider, \sysname's use of local PCs reduces the likelihood of hallucination while enhancing user control and understanding of the latent space, providing an effective approach to interacting with generative AI models.

\section{Related Work}

\subsection{Generative Models and Latent Spaces}

Understanding and manipulating latent spaces is central to controllable image generation. In this section, we review key techniques and differences in latent space structure across generative adversarial networks (GANs) and diffusion models.

StyleGAN2\cite{karras2020analyzing} is designed to generate high-quality, realistic images. 
It improves upon its predecessor, StyleGAN, by addressing issues like characteristic artifacts and improving image fidelity. 
Key innovations include an improved generator architecture that separates style and content, a redesigned path length regularization for smoother latent space navigation, and a rebalanced loss function for more stable training. 
StyleGAN2 is widely used in applications like image synthesis, style transfer, and data augmentation.

GANSpace\cite{harkonen2020ganspace} is a technique for discovering interpretable directions in the latent space of pretrained GANs, particularly StyleGAN. 
By applying PCA to intermediate latent representations, GANSpace identifies directions that correspond to meaningful semantic edits---such as changing facial expression, age, or hairstyle---in generated images. 
These directions can be manipulated using sliders, enabling intuitive and controllable image editing without requiring model retraining. 
GANSpace demonstrates that GAN latent spaces are highly structured and can be navigated in a semantically meaningful way.

Latent spaces in diffusion models\cite{ho2020denoising, song2020denoising, rombach2022high} are generally less interpretable than those in GANs due to fundamental architectural differences. 
While GANs generate images directly from structured latent vectors, diffusion models produce images by progressively denoising random noise, making their latent representations harder to analyze or manipulate. 
Unlike GANs, where latent directions can be found corresponding to meaningful attributes (e.g., age or expression), diffusion models lack clearly defined semantic axes. 
Recent advancements such as latent diffusion models (e.g., Stable Diffusion\cite{rombach2022high}) have improved interpretability by operating in compressed latent spaces derived from pretrained autoencoders. 
Although these models offer better image quality, their latent directions remain less semantically aligned than in GANs. 
Ongoing research is exploring methods like concept decomposition\cite{gandikota2025sliderspace} to enhance interpretability, but diffusion model latent spaces remain relatively underexplored when compared to GAN.

\subsection{Dimensionality Reduction and Its Interpretability}
Traditional dimensionality reduction (DR) methods aim to reduce the number of variables in high-dimensional data while preserving its essential structure. 
Principal Component Analysis (PCA) is a linear method that projects data onto orthogonal axes of maximum variance. 
Multidimensional Scaling\cite{cox2000multidimensional} (MDS) maintains pairwise distances between points, while Isomap\cite{balasubramanian2002isomap} preserves geodesic distances on a manifold. 
These techniques differ in their focus---some prioritize global structure, others local neighborhood fidelity---and are widely used in data analysis, visualization, and preprocessing.
t-SNE\cite{van2008visualizing} and UMAP\cite{mcinnes2018umap} are nonlinear methods that preserve local structures and are commonly used for visualization. 

A number of visualization and visual analytics tools and systems have been developed to help scientists and analysts explore and understand high-dimensional data in 2D, for example by aggregating data along certain dimensions\cite{stahnke2015probing} or augmenting dimensionality reduction plots with semantically meaningful gradients\cite{faust2018dimreader}.
DimBridge\cite{montambault2024dimbridge} enables users to directly interact with visual patterns in DR projections and retrieve corresponding data patterns from the original high-dimensional space. 
It employs first-order predicate logic to identify and present subspaces relevant to the selected visual patterns, thus facilitating deeper insight into the structure of the data. 


\subsection{Inverse Projection}
Inverse projection presents a greater challenge than direct projection because it requires reconstructing high-dimensional data from compressed low-dimensional representations. 
This difficulty stems from the need for generalization beyond training data and accurately synthesizing high-dimensional vectors from lower-dimensional inputs. 
Initial efforts to address this problem leveraged autoencoders, which jointly learn both the projection $P$ and its inverse $P^{-1}$, aiming to minimize reconstruction loss when mapping from $\R^n$ to $\R^m$, $n<m$\cite{hinton2006reducing}. 
However, despite their flexibility, autoencoders can be unintuitive to interpret\cite{van:2009:dim_reduction_survey} and often suffer from training instability\cite{vernier20}.

To provide more interpretable inverse mappings, Amorim et al. introduced iLAMP\cite{amorim:2012:ilamp}, which applies local affine transformations for reversing dimensionality reduction, based on earlier work from LAMP\cite{joia:2011:lamp}. 
Mamani et al. adopted a similar strategy for user-guided feature space adjustments\cite{mamani:2013:user_driven_feature_space_transformation}. 
iLAMP was later enhanced using radial basis functions (RBFs) to achieve smoother and more continuous inverse transformations, which proved effective in scenarios such as data augmentation\cite{amorim:2015:rbf}.

Beyond these methods, Kriegeskorte and Mur proposed an inverse approach to multidimensional scaling (MDS) by inferring dissimilarities from various 2D spatial arrangements\cite{kriegeskorte:2012:inverse_mds}. 
Cavallo et al. integrated inverse projection into Praxis\cite{cavallo:2018:praxis}, a visual analysis platform, combining the analytical inverse of PCA with autoencoder-based reconstruction. 
Similarly, Zhao et al.\cite{zhao:2020:chartseer} used a Grammar Variational Autoencoder (GVAE)\cite{kusner2017grammar} to enable interactive inverse projection of data charts, supporting visual steering and exploration tasks.

\subsection{Visualization Interfaces for Generative AI}

For graph layouts, Kwon et al.\cite{kwon2019deep}  propose a deep generative model using an encoder-decoder architecture, allowing intuitive exploration through a 2D latent space. 
The model offers a WYSIWYG interface, enabling users without technical expertise to generate layouts. 
This simplifies UI design for generative models by streamlining layout generation and improving visualization. 
However, the method struggles to generalize to graph structures significantly different from the training data. 
The quality of generated layouts depends on the dataset’s diversity and representativeness, making careful dataset selection essential.

UnProjection\cite{espadoto2021unprojection} learns an inverse-projection of a 2D dimensionality reduction projection back to high-dimensional data, supporting interactive exploration, interpolation, and synthesis. For generative model UIs, it allows users to navigate and manipulate outputs on the 2D plot. The inverse projection also enabled features such as decision boundary visualization, and gradient maps to enhance explainability and debugging. 
However, the method has limitations in its ability to maintain accuracy when dealing with complex or noisy data. 
While it provides a means to reconstruct high-dimensional data from 2D projections, the results can become unreliable when the data has high intrinsic dimensionality. 

Similarly, HyperNP\cite{appleby2022hypernp} uses neural networks to facilitate real-time, interactive exploration of projection hyperparameters, offering responsive controls and predictive projections for dimensionality reduction visualizations.
While offering real-time interactive exploration of projection hyperparameters, the method may experience reduced accuracy when hyperparameter gaps are large or training data is insufficient. 
As the number of hyperparameters increases, the training time can grow exponentially, making it challenging to scale the system to more complex projection models. 
Additionally, the smooth visual transitions generated by HyperNP might create a misleading sense of continuity, leading users to misinterpret the underlying changes. 

Ross et al.\cite{ross2021evaluating} suggested assessing the interpretability of generative models through tasks where users interactively adjust representation dimensions to recreate specific target instances.
GANSlider\cite{dang2022ganslider} investigated the use of feedforward visualizations, that is, previews of generated images, and their impact on users’ ability to control generative models.

For diffusion models, SliderSpace\cite{gandikota2025sliderspace} enables intuitive model control through sliders that adjust visual attributes. 
It allows users to create outputs via decomposed concepts, and gain insights into model behavior, all without technical expertise.
The SliderSpace framework, though effective in providing intuitive control over diffusion models, has limitations in the interpretability of slider directions as these sliders depend on the quality of the concept decomposition, and users may struggle to predict how adjusting a slider will impact the visual output. 

Our system builds on prior efforts to make generative models more accessible and interpretable, such as GANSlider and SliderSpace, by emphasizing intuitive control through low-dimensional interactions. 
Unlike methods like Deep Generative Graph Layouts\cite{kwon2019deep} or UnProjection\cite{espadoto2021unprojection}, which are limited by model generalization or the reliability of inverse projections, our proposed approach adapts locally to the latent space, ensuring more accurate reconstructions even in high-dimensional settings.
\section{Problem Statement and Motivation}


\begin{figure}[t]
    \centering
    \begin{subfigure}[t]{0.49\linewidth}
        \includegraphics[width=\linewidth]{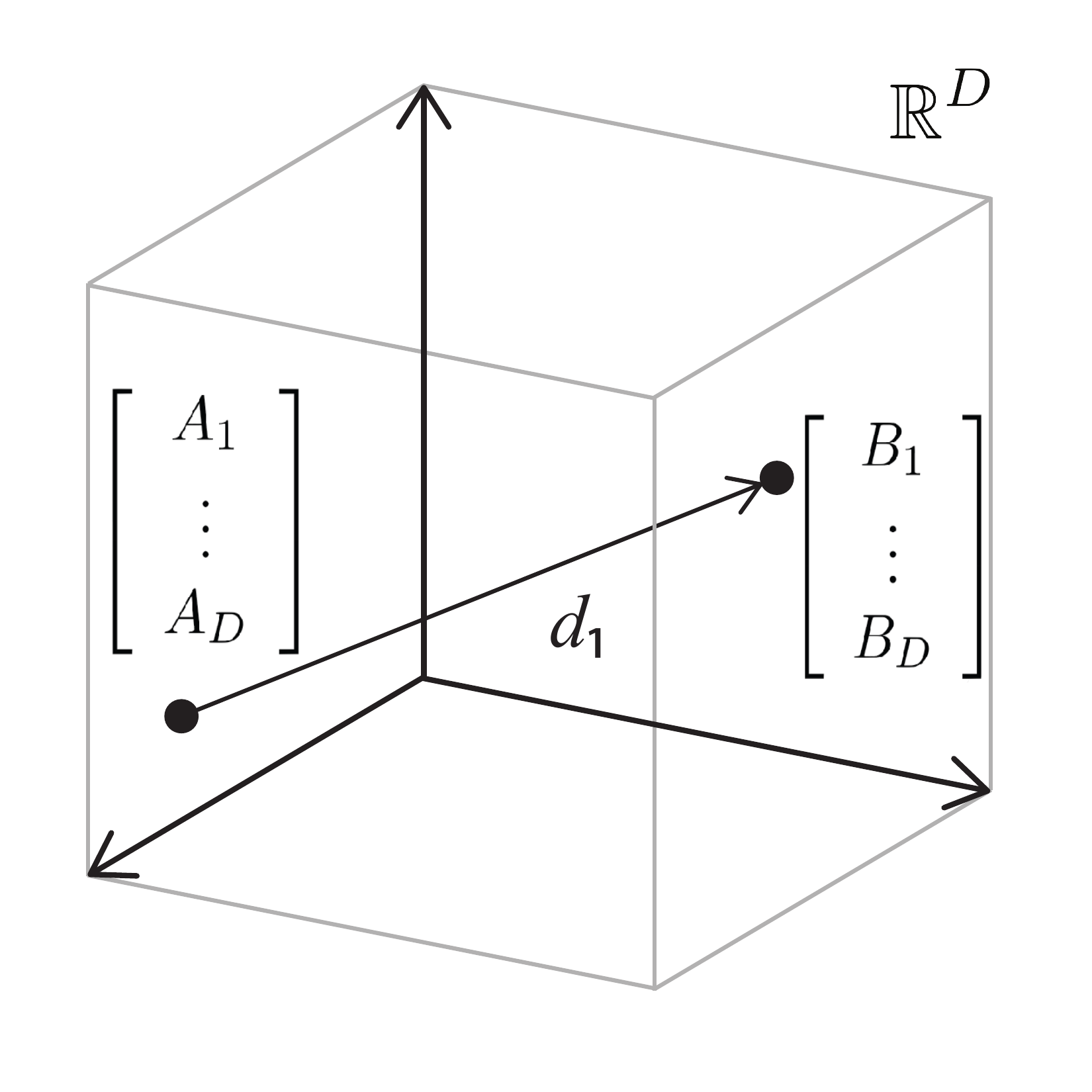}
        \caption{Line formed in 1D between two data points in $\R^D$}
        \label{fig:illu-proj-1}
    \end{subfigure}
    \begin{subfigure}[t]{0.49\linewidth}
        \includegraphics[width=\linewidth]{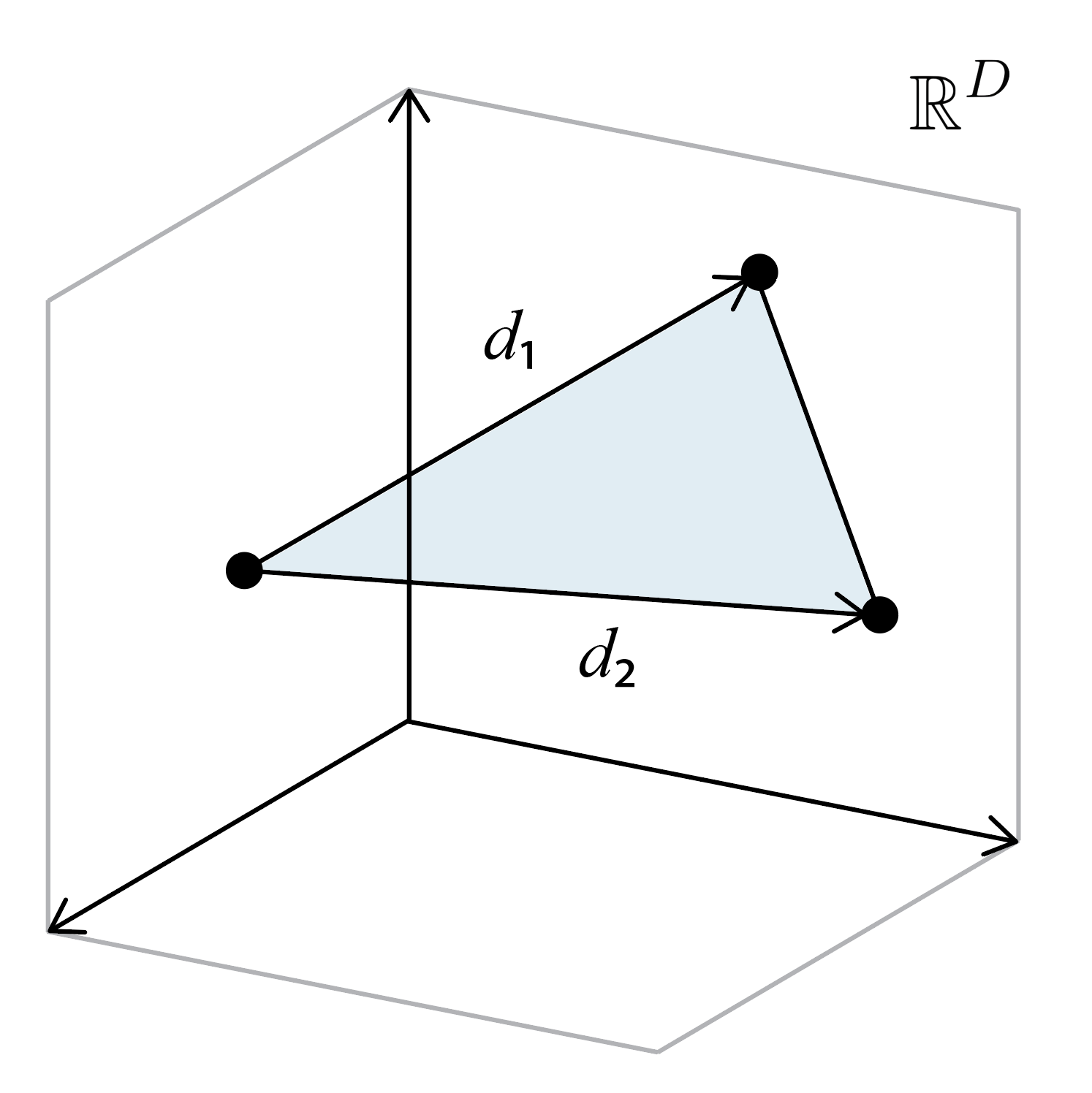}
        \caption{A 2D plane formed between three data points in $\R^D$}
        \label{fig:illu-proj-2}
    \end{subfigure}
    \begin{subfigure}[t]{0.49\linewidth}
        \includegraphics[width=\linewidth]{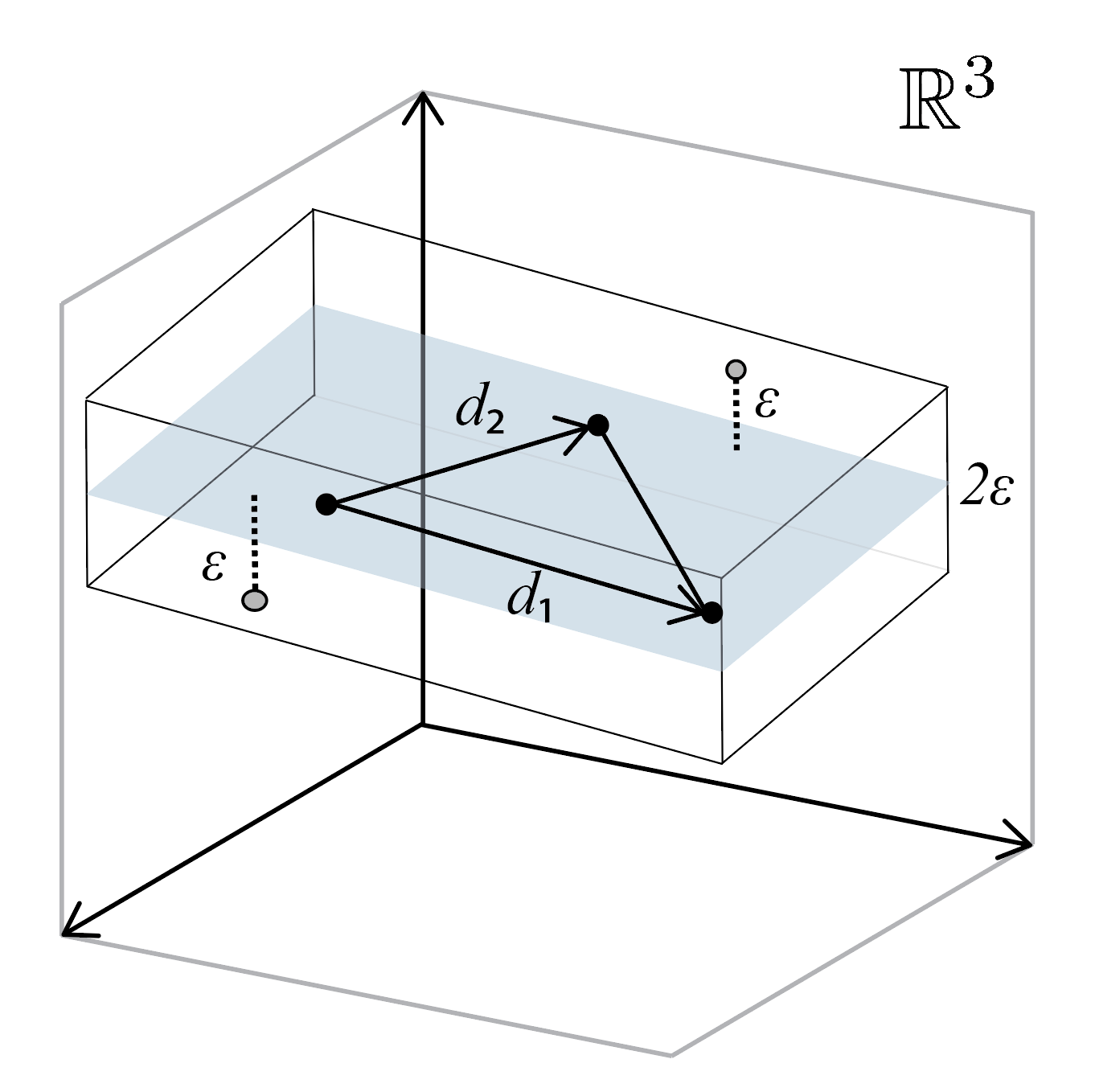}
        \caption{A 2D plane in $\R^3$ with data points within $\epsilon$ projected}
        \label{fig:illu-proj-3}
    \end{subfigure}
    \begin{subfigure}[t]{0.49\linewidth}
        \includegraphics[width=\linewidth]{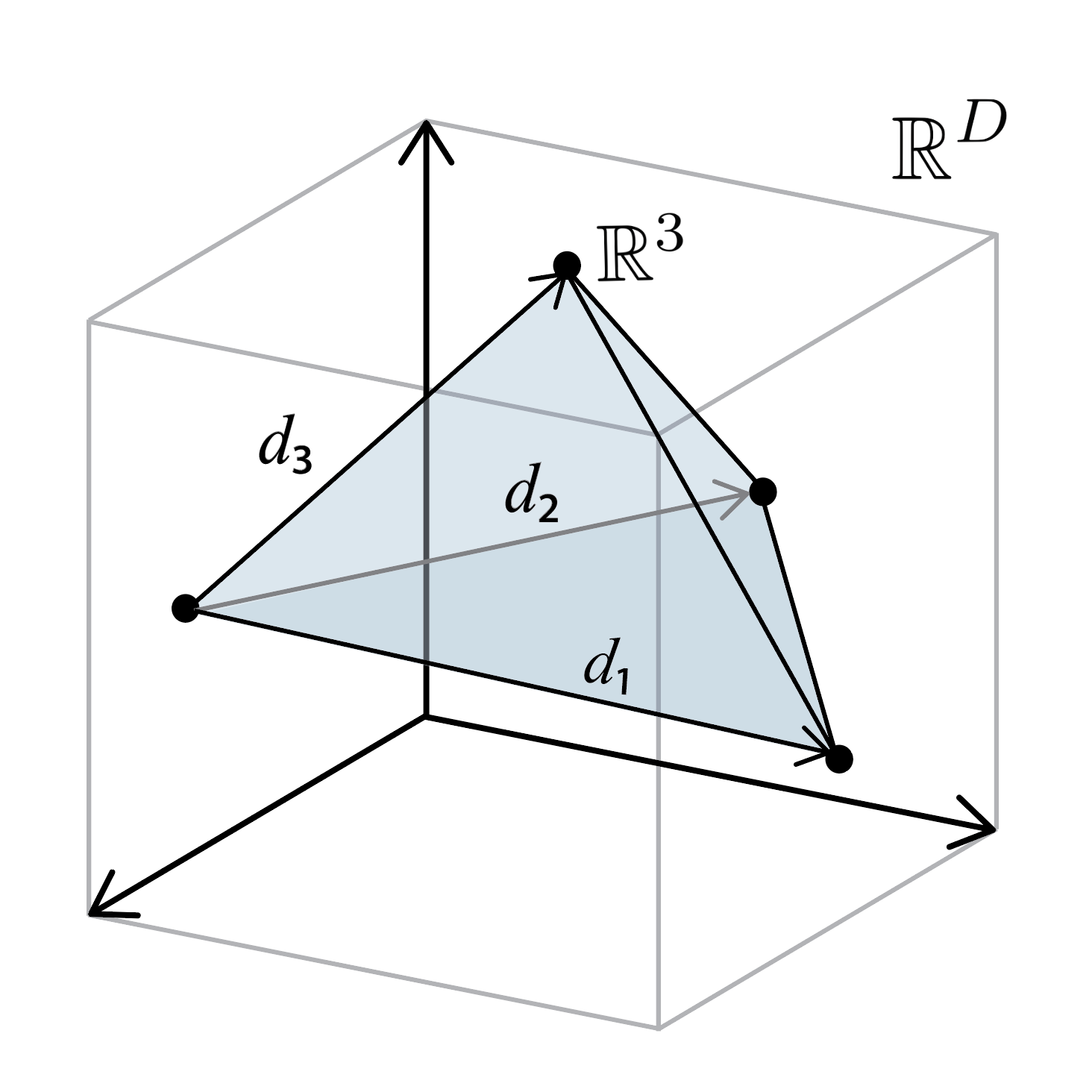}
        \caption{Adding another dimension forms a 3D volume in $\R^D$ 
        }
        \label{fig:illu-proj-4}
    \end{subfigure}
    \caption{Intuition behind our approach. (a) We begin with the observation that interpolation between two points in $D$-dimensional space traces out a 1D line segment. (b) This idea naturally extends to a 2D plane when interpolating among three points. (c) By allowing a small error tolerance $\epsilon$, additional points can be approximated within the same 2D plane, forming a locally linear neighborhood. (d) Finally, we relax the 2D constraint and generalize this approach to higher-dimensional subspaces within $\R^D$ to accommodate more complex local structures.}
\end{figure}
Visualization is poised to play a critical role in human-AI interaction, offering users more precise and expressive control over AI systems than modalities such as speech. In addition to enabling high-precision input, visualizations also provide interpretable feedback beyond merely presenting the model’s outputs.

However, a core challenge in visualization lies in its intrinsic limitation: the dimensionality of user input and system output is typically constrained to two dimensions on a screen. As illustrated in Equation~\ref{eq:GenAI-VIS}, extending interactions from 2D to a $D$-dimensional latent space is nontrivial, as 2D inherently encodes less information than a $D$-dimensional space.

Inverse projections techniques, such as iLAMP\cite{amorim:2012:ilamp}, Praxis\cite{cavallo:2018:praxis}, and UnProjection\cite{espadoto2021unprojection}, attempt to address this by learning the inverse of a dimensionality reduction mapping (i.e., approximating $\text{GenAI-VIS} := \text{DR}^{-1}$, where DR denotes a dimensionality reduction function in Equation~\ref{eq:DR}). 
These techniques rely on the assumption that the structure of the original $D$-dimensional data can be meaningfully captured in 2D. 
For relatively simple datasets like MNIST, CIFAR-10, or Fashion-MNIST, this assumption can hold, making inverse projection techniques effective. 
However, for more complex data---particularly those involving high-dimensional generative models---this assumption breaks down. 
The projection of rich, high-dimensional structures into 2D can result in substantial information loss that inverse techniques cannot recover.

Our approach to solving Equation~\ref{eq:GenAI-VIS} diverges from these inverse-projection techniques. We begin with a simple yet practical observation: in the limit, generating a high-dimensional vector can be unambiguously reduced to a one-dimensional interaction. 
For instance, interpolating between two points in $D$-dimensional space follows a linear path--a fundamental 1D structure (see Figure~\ref{fig:illu-proj-1}). Extending this idea, interpolation among three points lies on a 2D plane, which still conforms to the constraints of a visualization on a 2D screen  (see Figure~\ref{fig:illu-proj-2}). 
By allowing a small tolerance, $\epsilon$, points that do not lie perfectly on the 2D plane can also be represented, resulting in a compact 2D representation of a localized high-dimensional manifold  (see Figure~\ref{fig:illu-proj-3}. We use here a 3D data space to better illustrate the concept).

Finally, we further relax the 2D constraint to include multiple linear dimensions (see Figure~\ref{fig:illu-proj-4}).
Expanding the subspace in this way allows more data points to be represented, forming a locally linear neighborhood within the high-dimensional space. 
While this approach exceeds the limitations of a 2D display, we assume users can interact with multiple control elements (such as more than two sliders or coordinated scatterplots) to explore these higher-dimensional neighborhoods. 

This introduces a design tradeoff between the \textbf{local dimensionality} and the \textbf{number of local neighborhoods}. Restricting the local dimensionality reduces the number of control elements (e.g., fewer sliders), but increases the number of neighborhoods needed to sufficiently cover the latent space. Conversely, increasing local dimensionality allows for broader coverage with fewer neighborhoods, but requires more control elements (e.g., sliders) for a user to navigate each one.

Guided by this intuition, we introduce \sysname. In the following section, we present: (1) a technique for estimating the intrinsic dimensionality of low-dimensional manifolds embedded in high-dimensional latent spaces, and (2) a method for automatically identifying local neighborhoods using this estimation technique.

\section{Method}
In this section, we present our methodology for analyzing high-dimensional datasets by leveraging Multiscale Singular Value Decomposition (MSVD)\cite{little2009estimation} and local Principal Component Analysis (PCA). 
MSVD estimates intrinsic dimensionality by examining geometric properties at multiple scales, effectively distinguishing data structure from noise and curvature effects. 
Using the MSVD method to identify local neighborhoods, 
we construct a connectivity graph that captures local relationships while preserving the dataset’s global structure, visualized using UMAP or a force-directed layout for clarity. 
To further explore local variations, we propose three visualization interfaces, including PCA scatterplot, filmstrip preview and image grid previews that allow a user to explore along principal component directions, enabling the synthesis of new images that reveal continuous transformations in the data. 
Our overview + detail approach provides a structured, interpretable, and interactive means of understanding complex latent dimensions in the generative model.

\subsection{Multiscale SVD}


We apply MSVD to estimate the intrinsic dimensionality of noisy, low-dimensional manifolds embedded in a high-dimensional space. 
This method provides a robust way to determine the intrinsic dimension of a data manifold, even when it is affected by noise and curvature. 
Given a set of noisy samples $X$ concentrated around a Riemannian $k$-manifold $(M, g)$ within a $D$-dimensional ambient space, the goal is to estimate the intrinsic dimensionality $k$, which represents the true degrees of freedom of the data structure.

The algorithm operates by defining local neighborhoods around each data point. Specifically, for every point $z \in X$, we construct a local neighborhood $X(z, r) = X \cap B_z(r)$, where $B_z(r)$ is a ball of radius $r$ centered at $z$. These neighborhoods capture local geometric properties while allowing the analysis to scale dynamically across different regions of the manifold.

Within each neighborhood, the algorithm computes the covariance matrix $\text{cov}(X(z, r))$, which encodes the spread of data points within that region. By performing Singular Value Decomposition (SVD), we extract its singular values $\{\sigma_i^{(z,r)}\}_{i=1}^{D}$, which provide information about the local structure of the data. 
Since different singular values correspond to different sources of variation, the algorithm analyzes their behavior across multiple scales $r$. 

\begin{figure}
    \centering
    \includegraphics[width=1\linewidth]{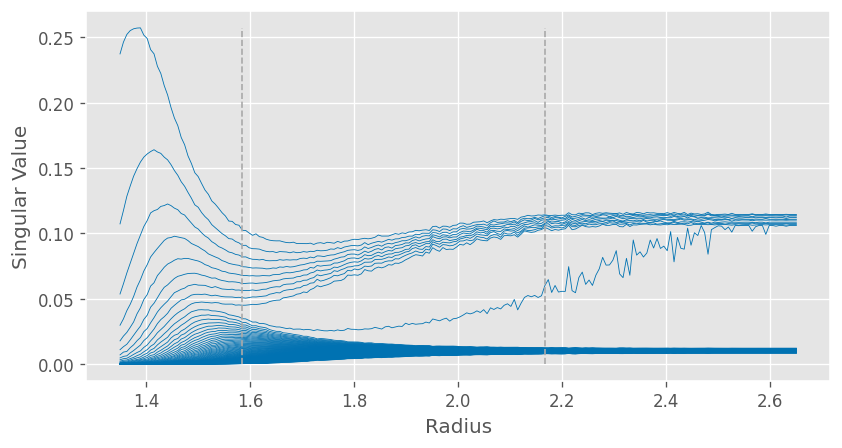}
    \caption{
    Examples of singular value spectra as a function of scale, as illustrated in\cite{little2009estimation}, show data occupying $S^9$ within a 100-dimensional ambient space. The top nine singular values correspond to the intrinsic dimensions, the tenth reflects curvature effects, and the remaining 90 capture noise in the data. Vertical lines show the optimal scale range where noise levels remain low and curvature-induced distortions are not as dominant.
    }
    \label{fig:illu-msvd}
\end{figure}

As illustrated in Figure~\ref{fig:illu-msvd}, the eigenvalue spectrum reveals distinct patterns corresponding to different sources of variation. 
The dominant eigenvalues, associated with the intrinsic dimensions of the data manifold, exhibit linear growth with increasing scale. 
In contrast, curvature effects manifest as eigenvalues that grow quadratically, while noise-related eigenvalues remain relatively stable across scales. 

The separation between the eigenvalues enables the algorithm to differentiate between intrinsic structure, curvature distortions, and noise, ensuring a robust estimation of the intrinsic dimensionality.
By examining how these eigenvalues change across different values of $r$, the algorithm automatically identifies the optimal scale ranges for estimating noise levels and detecting curvature-induced distortions, and provides an adaptive way to determine intrinsic dimensionality $k$.

Unlike global PCA, which operates on global properties of the data, MSVD operates locally, making it effective in analyzing complex datasets. 
The ability to separate intrinsic geometry from noise makes it valuable for applications in manifold learning, data visualization, and scientific data analysis, where understanding the underlying structure of high-dimensional data is crucial. 


To estimate the intrinsic dimensionality, MSVD first estimates an upper bound for the intrinsic dimension.
To simplify the analysis, we omit the individual variance among different center points and calculate the average singular values across all sample data points as $\sigma_i^{(r)} = \frac{1}{N} \sum_{z = 1,2,...N} \sigma_i^{(z,r)}$.
One can also do the same dimension estimate at each point.
MSVD identifies manifold dimension upper bound by detecting linearly increasing singular values at larger scales. 
As shown in Figure~\ref{fig:illu-msvd}, the singular values corresponding to intrinsic dimensions tend to increase with scale, whereas those representing noise dimensions typically decrease. 
MSVD then estimates the number of dimensions whose singular values increase at a rate that surpasses a predefined threshold.
Let $m_i$ represent the slope estimate for the $i^{th}$ singular value $\sigma_i^{(r)}$ over the large-scale range, defined as $r \in [r_{large}, r_{max}]$. 
Here, large scales are characterized by singular values that exceed $50\%$ of the max scale, where max scale estimates the radius of the point cloud: the minimum distance from a set of sampled points to the furthest points in the dataset 
i.e.: 
\begin{equation}
{r_{max} = min \{ max \{\text{distance}(X_z,\; X)\} \text{ for } z \in \text{ sample } \} } 
\end{equation}

A dimension is considered a signal if its slope is greater than a fraction of its maximum singular value per unit scale. Specifically, the $i^{th}$ dimension is classified as a signal if:
$$m_i > \epsilon \frac{\sigma_i^{(r_{max})}}{r_{max}}$$
Otherwise, it is regarded as noise. 
This approach allows MSVD to effectively differentiate between intrinsic and noise dimensions, providing a reliable estimate of the upper bound for the intrinsic dimension.

Next, among the signal dimensions, MSVD eliminates those that grow quadratically by performing a quadratic fit to the singular values as a function of the scale. A dimension is considered to exhibit quadratic growth if the coefficient for the quadratic term is significantly greater than zero, indicating that the singular values are increasing at a quadratic rate. This step further refines the dimensionality estimate by excluding artificially inflated dimensions due to data curvature, leading to a more accurate representation of the underlying data manifold.


\subsection{Local PCA and Connectivity Graph Layout}

\begin{figure}[t]
    \centering
    \includegraphics[width=0.32\linewidth]{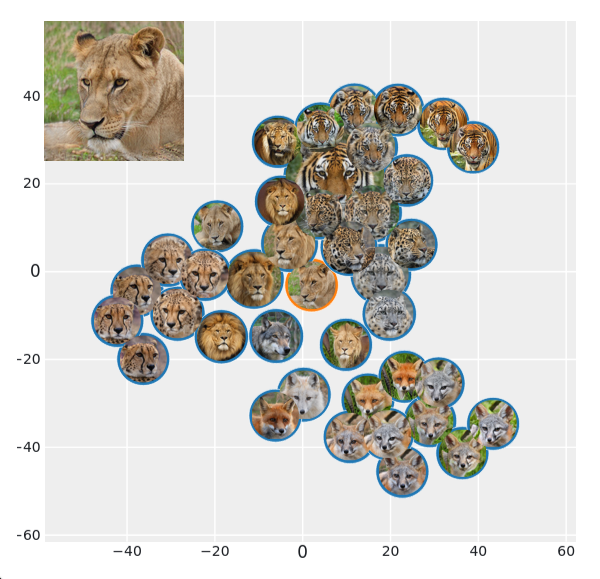}
    \includegraphics[width=0.32\linewidth]{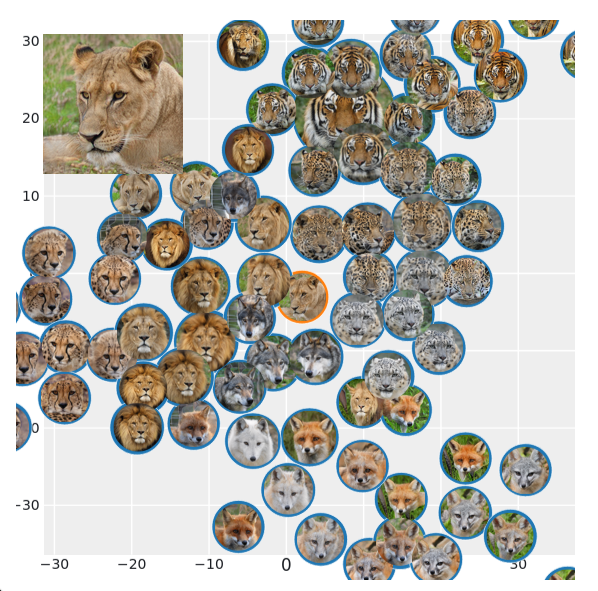}
    \includegraphics[width=0.31\linewidth]{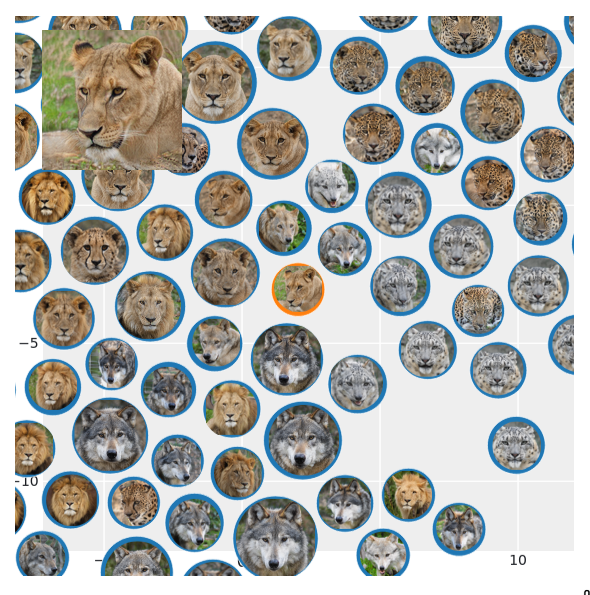}
    \caption{Graph layout of the latent space with semantic zooming.}
    \label{fig:zoomable-layout}
\end{figure}




After estimating the intrinsic dimension, we construct a neighborhood graph using the optimal radius determined by MSVD 
to represent the structure of the dataset. 
This graph-based representation helps enclose local relationships while preserving the global connectivity of the data. 
Each node in the graph corresponds to a subset of data points within a local neighborhood, with center points selected randomly and iteratively to ensure comprehensive coverage of the dataset. 
Using the largest optimal scale $r$ determined by MSVD, we iteratively generate balls of radius $r$ centered on randomly chosen points. 
Any sample point already enclosed by an existing ball is excluded from the selection pool for new centers.
This iterative selection process prevents gaps in the representation and ensures that all data regions are adequately accounted for.

Within each local neighborhood enclosed by the balls, we apply PCA  
to identify the dominant directions of variation and perform dimensionality reduction. 
The relationships between nodes are encoded as edges, where edges exist if two nodes share overlapping data points. 
This connectivity structure enables a smooth transition between different neighborhoods and provides insight into how local structures relate to the overall data manifold.

While the data samples can be shown as points in a scatter plot using dimensionality reduction techniques such as UMAP (Figure~\ref{fig:latentgandr-0} (top left)),
to facilitate an intuitive and meaningful visualization of the graph, we employ a force-directed layout, where nodes experience repelling forces while edges exert attractive forces, resulting in an organic arrangement that reflects underlying data relationships (Figure~\ref{fig:teaser} (left)),
This layout helps preserve local structures while avoiding excessive clutter or distortion in the visualization.

To further enhance clarity and usability, we incorporate collision detection and removal techniques to minimize node overlap, ensuring that individual nodes remain distinguishable. 
Additionally, an interactive interface integrates semantic zooming, allowing users to explore the graph at multiple levels of detail. 
Figure~\ref{fig:zoomable-layout} shows this graph layout with semantic zooming.
This feature enables both high-level overviews and fine-grained inspections of local data structures.

In the final presentation, we hide the edges to reduce visual clutter and enhance readability, focusing on the spatial organization of nodes. 
To provide additional context for each node, we generate a thumbnail image derived from the average style vector of its associated data points. 
These thumbnails serve as a compact visual summary of each neighborhood, offering an intuitive way to interpret patterns and variations across different regions of the dataset. 
This approach results in a structured, interpretable, and interactive representation of the data manifold, aiding in both analysis and exploration.

\subsection{Visualizing Local PCA Slices}

\begin{figure}[t]
    \centering
    \includegraphics[width=1\linewidth]{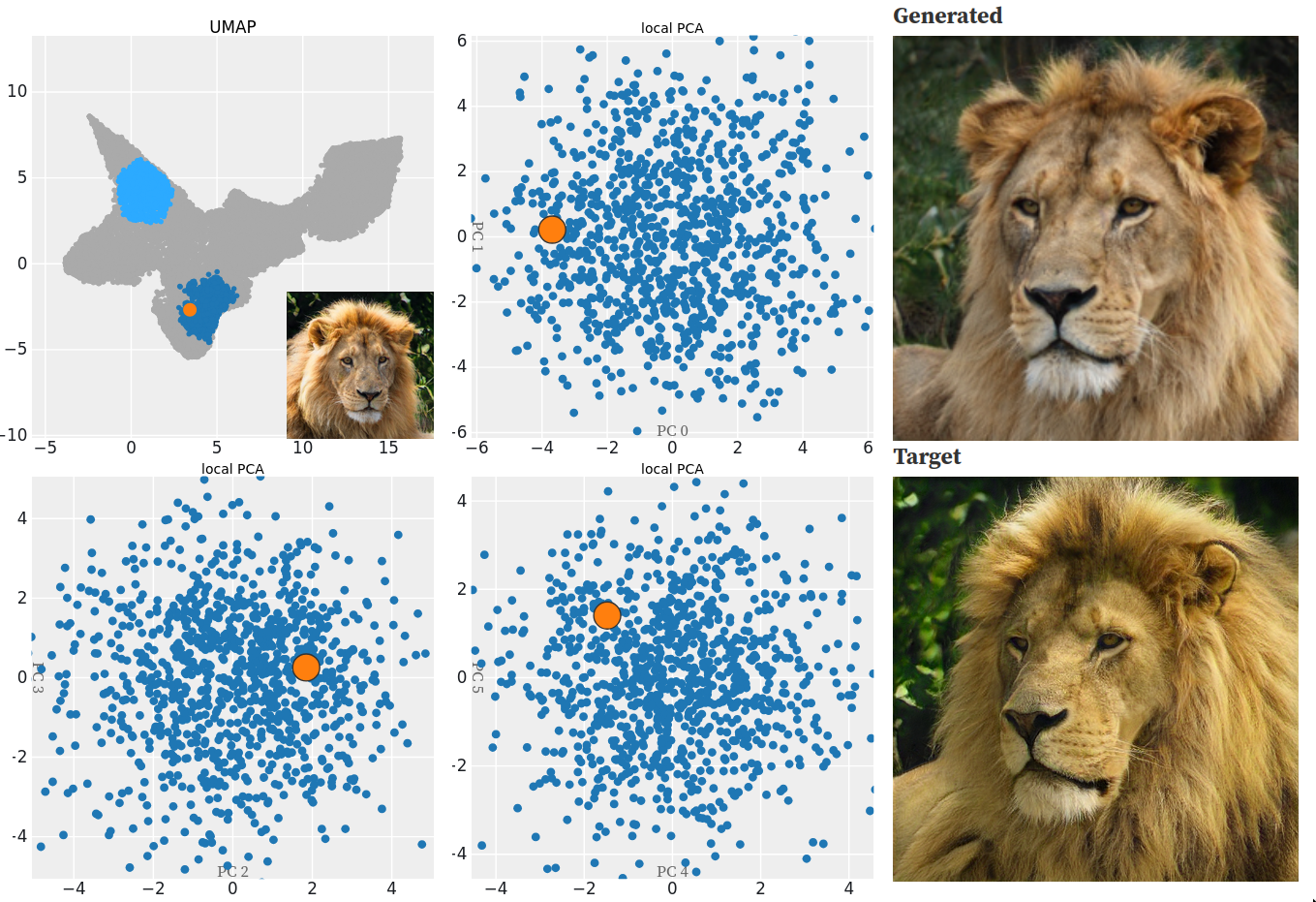}
    \caption{Local PCA scatterplots in \sysname.}
    \label{fig:latentgandr-0}
\end{figure}

\begin{figure}[t]
    \centering
    \includegraphics[width=1\linewidth]{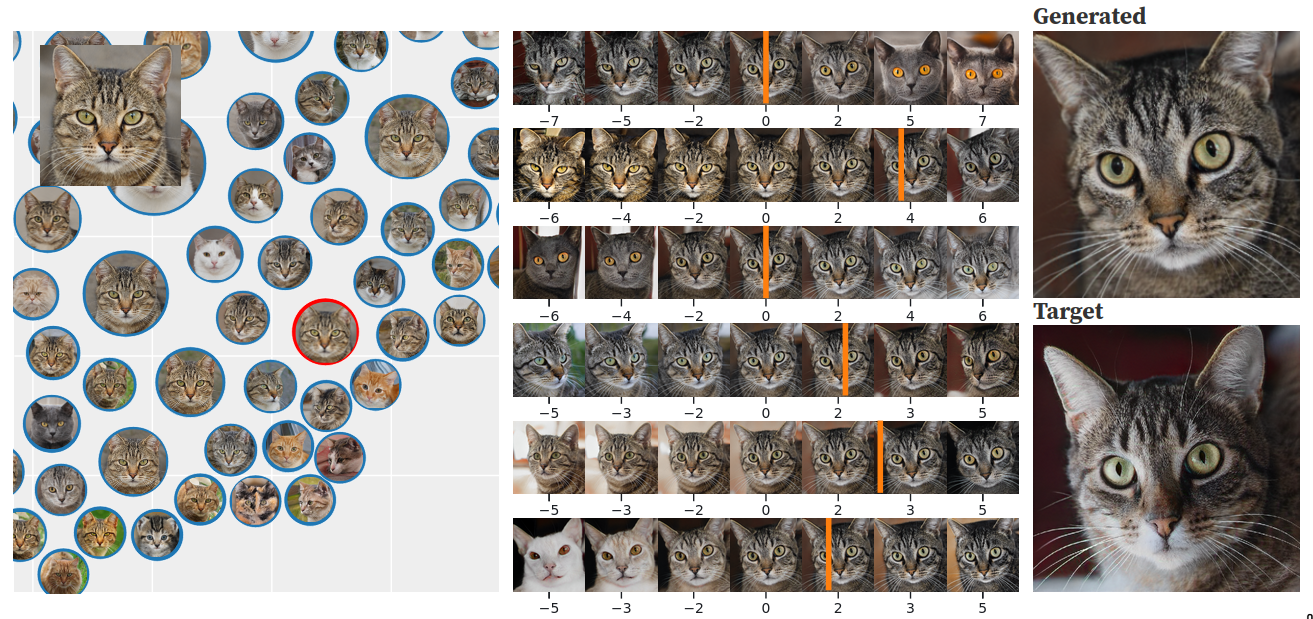}
    \caption{\sysname that incorporates GANSlider which shows local PCA as with filmstrip sliders. Using AFHQ cat dataset as an example.}
    \label{fig:ui-latentgandr-slider}
\end{figure}

To analyze the local structure of the data, we visualize local structures of generated data through three methods: (1) PCA scatterplots, (2) filmstrip-like previews used in GANSlider, or (3) 2-dimensional image grids along principal components (PC). 
These techniques provide a structured way to examine variations within the dataset by leveraging 
PCA to identify dominant axes of local variation. 

\smallskip
\noindent\textbf{PCA scatterplots:} As the most straightforward way to analyze local structure of a data neighborhood, we project points in the neighborhood with PCA. 
Figure~\ref{fig:latentgandr-0} shows a neighborhood of lions projected onto pairs of consecutive principal components. 
Although this visualization technique does not offer previews like the other methods, it is the most robust and generalizable across diverse data types, including text, documents, and audio.

In the later two cases, instead of simply reducing dimensionality and projecting existing data points, we use PCs as bases to generate new images along the PC directions, enabling a detailed preview of how the dataset variations.

\smallskip
\noindent\textbf{Filmstrip Preview:} We adopt the approach used in GANSlider, which samples data along each principal component at fixed step sizes to generate images. In our case, however, PCA is applied locally within a neighborhood, capturing only the local variance of the latent space. Figure~\ref{fig:ui-latentgandr-slider} illustrates an example of this interface.

\smallskip
\noindent\textbf{Image Grids:} 
For each pair of consecutive principal components, we construct a two-dimensional image grid that visualizes synthesized data points. These grids systematically sample data along the principal axes, capturing not only the semantic meaning encoded in each direction but also the stylistic interactions between adjacent components. See Figure~\ref{fig:teaser} for an example.


All visualizations are interactive, allowing viewers to hover over or click on scatterplots, sliders, or image grids to view AI-generated images. 
These interactions enable a viewer's continuous exploration of transformations between data points, revealing the underlying patterns in the data such as variations in style, shape, orientation, or texture. 
\section{Evaluation}

Quantitatively, we conduct one internal and two external evaluations to assess the accuracy and effectiveness of \sysname. 

For internal validation, we analyze the eigenvalue spectrum obtained from MSVD to examine the separation between dominant and non-dominant dimensions. 
A distinct gap in the eigenvalue distribution serves as an indicator of well-defined intrinsic dimensionality, distinguishing meaningful data structure from noise and extrinsic curvature effects. 
By systematically assessing these eigenvalue trends across different scales and neighborhoods, we confirm the robustness of our estimation method.

For external validations, we first compare the reconstruction error of data points under global and local PCA. 
This measures how much points deviate from their low-dimensional projections, allowing us to assess the distortion introduced by Local PCA (used by \sysname) versus the Global PCA approach employed by GANSlider.

Second, we compare the pairwise distance distributions among points in a local neighborhood across three dimensionality reduction techniques: Global PCA, UMAP, and our proposed Local PCA. This analysis reflects the quality of the local neighborhoods---those with smaller pairwise Euclidean and geodesic distances better preserve the underlying local geometry.



\subsection{Internal Validation}
\begin{figure}[t]
    \centering
    \includegraphics[width=1\linewidth]{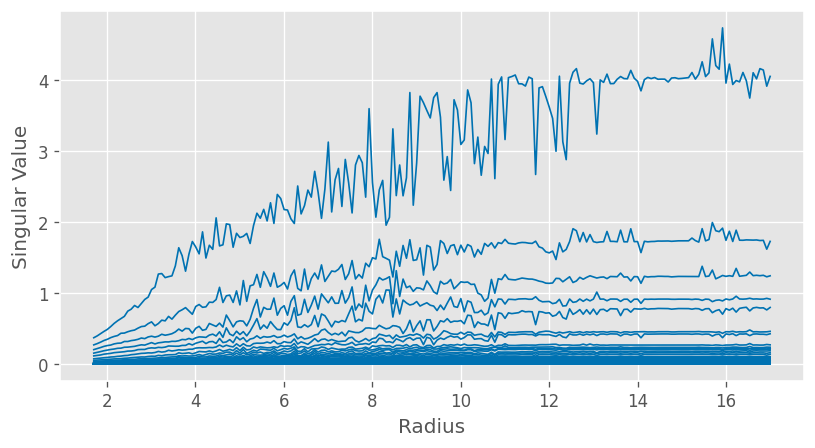}
    \caption{
    Singular value spectra as a function of scale for the AFHQ wild animal dataset, with an intrinsic dimension estimate of 5. Vertical lines indicate the optimal scale range where noise levels remain low, and curvature-induced distortions are minimal. No significant curvature was observed among the dimensions.
    }
    \label{fig:eval-msvd}
\end{figure}


To validate the intrinsic dimension estimation, we analyze the separation between dominant and non-dominant dimensions by visualizing the eigenvalue spectrum obtained from MSVD. 
Specifically, we plot the eigenvalues as a function of scale to observe the characteristic gap that emerges between the significant dimensions, which correspond to the intrinsic structure of the data, and the remaining dimensions, which primarily capture noise or extrinsic curvature effects. 
A clear and stable gap indicates a well-defined intrinsic dimension, while its absence suggests ambiguity or challenges in the estimation process. 
By systematically examining these eigenvalue distributions across multiple local neighborhoods, we ensure the robustness of our dimensionality assessment, verifying that the estimated intrinsic dimension accurately reflects the underlying geometric structure of the data.

Figure~\ref{fig:eval-msvd} shows the eigenvalue spectrum of the AFHQ wild animal dataset, with a distinct gap between dominant and residual dimensions, validating the intrinsic dimensionality estimation.
The dominant eigenvalues correspond to meaningful variations within the data manifold, while the smaller eigenvalues capture noise or extrinsic curvature effects. 
As the scale parameter increases, we observe that the eigenvalues associated with the intrinsic dimensions grow linearly, whereas those influenced by curvature exhibit quadratic growth. 
This pattern aligns with theoretical expectations, confirming that our method effectively distinguishes intrinsic structure from artifacts introduced by noise and embedding distortions. 
The visualization provides an intuitive way to assess dimensionality, ensuring that the chosen intrinsic dimension reflects the true underlying degrees of freedom in the dataset.

\subsection{External Comparison}

\begin{figure}[t]
    \centering
    \includegraphics[width=0.99\linewidth]{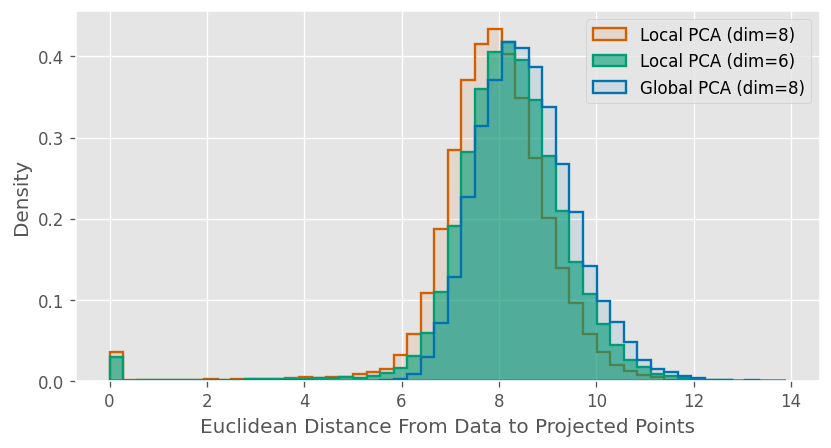}
    \caption{Distribution of distances (the lower the better) from data points to their projections with different projection methods using the AFHQ Wild Animals dataset.}
    \label{fig:eval-dist-to-projection}
\end{figure}

\begin{figure}[ht]
    \centering
    \includegraphics[width=0.99\linewidth]{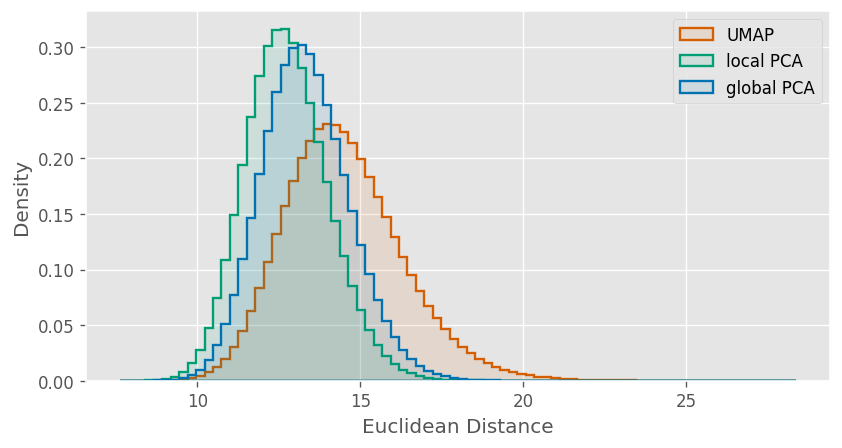}
    \includegraphics[width=0.99\linewidth]{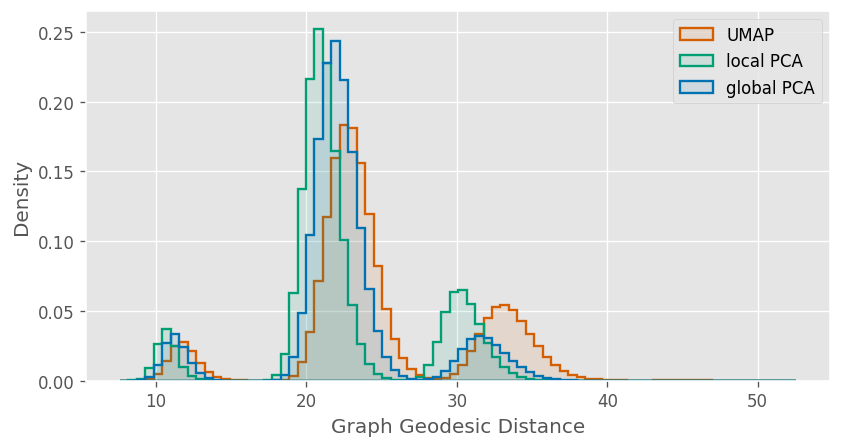}
    \caption{Distribution of pairwise distances between data points in a local neighborhood using UMAP, Global PCA, and Local PCA. With both: (top) Euclidean and (bottom) geodesic distances, Local PCA outperforms the other two with distributions centered at the smallest values, indicating superior locality preservation.}
    \label{fig:eval-pdist}
    \vspace{-0.5cm}
\end{figure}

\subsubsection{Measuring Reconstruction Error}
To assess the effectiveness of our Local PCA approach, we compare the reconstruction error of data points under both global and local PCA. 
This metric quantifies how much each point deviates from its low-dimensional projection, providing insight into how well each method preserves the intrinsic structure of the data. 
Lower reconstruction errors indicate more faithful representations and reduced information loss. 

We analyze the distribution of these errors across all data points to detect patterns of distortion or bias. 
As shown in Figure~\ref{fig:eval-dist-to-projection}, Local PCA consistently achieves lower distortion than Global PCA. 
Notably, Local PCA with just six dimensions captures more local structure than Global PCA with eight dimensions, highlighting its superior ability to preserve local data geometry.

\subsubsection{Comparing Pairwise-Distance Distributions in Local Neighborhoods}

Next, we compare the distribution of pairwise distances between data points in a local neighborhood using three methods: Global PCA, UMAP, and our Local PCA. 
Global PCA projects data based on global data variance, often distorting local relationships by prioritizing large-scale structure. 
UMAP, while nonlinear and designed to preserve local neighborhoods, can introduce artifacts when embedding high-dimensional data into low-dimensional spaces. 
In contrast, our Local PCA method constructs adaptive neighborhoods and performs dimensionality reduction within them, yielding a more faithful representation of the local data manifold.

We analyze the pairwise distance distributions using both Euclidean and geodesic distances to evaluate how well each method preserves local geometry. 
Global PCA tends to inflate distances due to its rigid linear projection, resulting in a loss of fine-grained local structure. 
UMAP generally produces a more compact distribution in 2D, preserving local connectivity but often distorting actual distances due to the constraints of low-dimensional embedding. 
Local PCA, by comparison, better maintains the structure of pairwise distances, balancing neighborhood preservation with minimal distortion. 
As shown in Figure~\ref{fig:eval-pdist}, Local PCA yields the distribution centered at the smallest values with both Euclidean and geodesic distances, indicating superior locality preservation.

This improved local fidelity is particularly important for downstream tasks such as image synthesis and data visualization, where retaining accurate local structure is essential for generating meaningful and interpretable outputs.

\section{Pilot Study}

We conducted a controlled experiment to evaluate the effectiveness of LatentGandr compared to GANSlider (Filmstrip) in a user study. The experiment assesses user interaction patterns, task performance (image reconstruction accuracy), and collects subjective feedback to determine the advantages of different latent space exploration interfaces.

\begin{figure}[t]
    \centering
    \includegraphics[width=1\linewidth]{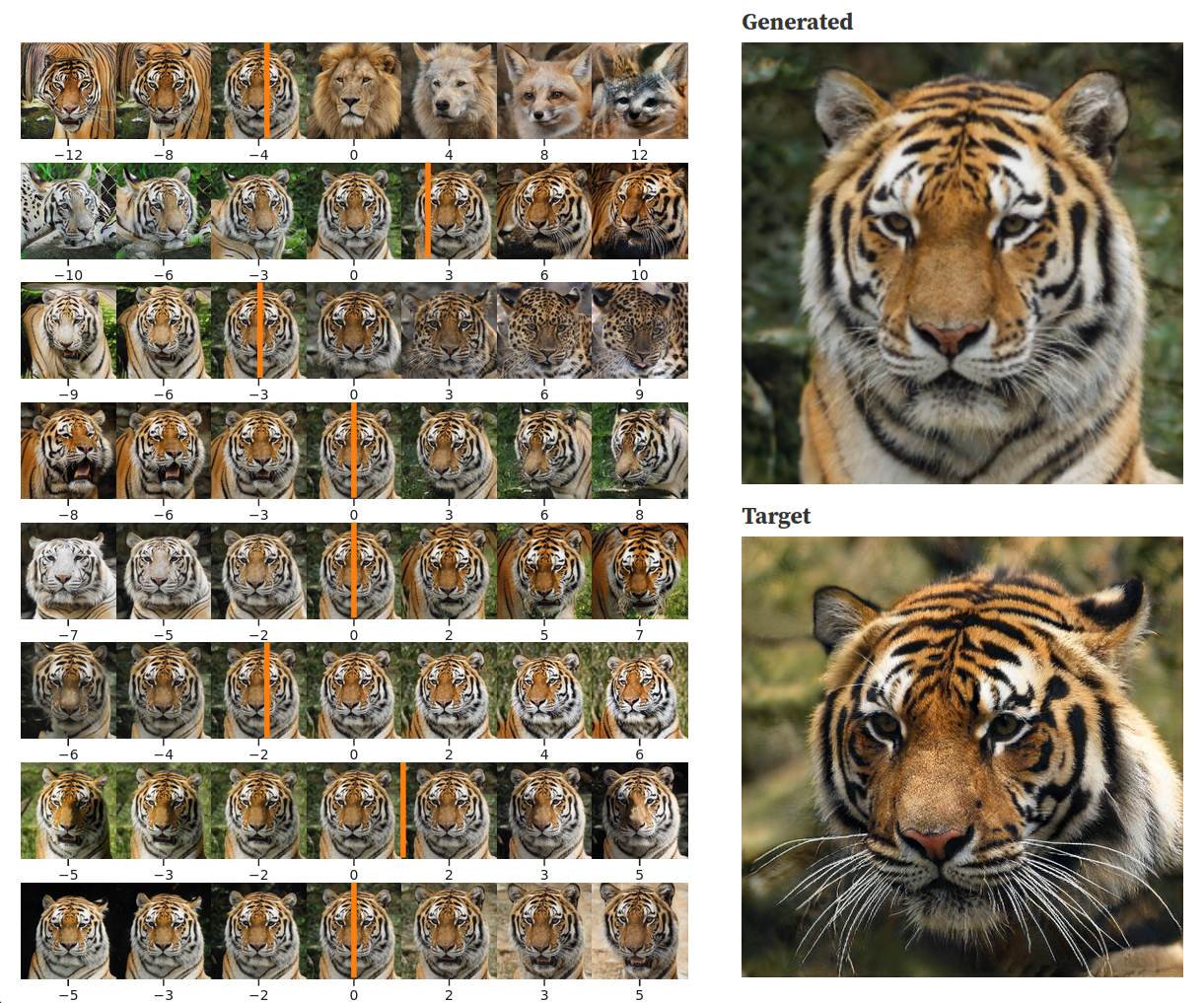}
    \caption{GANSlider interface used in the study.}
    \label{fig:ui-gs}
\end{figure}

\subsection{Study Design}
\textbf{A/B Testing:}
We employed an A/B testing approach with two interface conditions. The baseline condition, GANSlider (Figure~\ref{fig:ui-gs}), utilized Global PCA with a slider-based interaction model. The variation condition, LatentGandr, combined a Local PCA approach with a two-part interface: a force-directed layout on the left for selecting local neighborhoods, and interactive image grids on the right for fine-tuning along locally extracted principal components (Figure~\ref{fig:teaser}). 

\smallskip\noindent{\textbf{Independent Variables:}}
The primary independent variable in our study was the tool type, where all participants completed both the Baseline (GANSlider) and Variation (LatentGandr) in random order. 

\smallskip\noindent{\textbf{Dependent Variables:}}
We measured both objective interaction metrics and subjective feedback. Interaction measures included the total number of interactions (clicks, movements, selections), completion time per task and per participant, and task success, quantified as the distance between the final reconstructed image and the target image. Subjective feedback was collected through NASA-TLX, Likert-scale ratings assessing ease of use, satisfaction, and perceived effectiveness, and open-ended responses for qualitative insights.

\smallskip\noindent{\textbf{Key Metrics:}}
A subset of the dependent variables was selected as key performance indicators. These included the total number of interactions, image reconstruction accuracy measured by the distance between the final and target image, and overall completion time.

\subsection{Apparatus}
\noindent{\textbf{Web System:}}
The study was conducted using a custom web-based system, with a front-end developed using the D3.js framework and a back-end implemented in Flask. The system was hosted on a server with a NVIDIA GeForce RTX 4090 GPU to ensure consistent performance in interaction and rendering. Target images were generated using the StyleGAN2 model. 

\smallskip\noindent{\textbf{Image Reconstruction Tasks:}}
Target images were generated using StyleGAN2. To control task difficulty and ensure consistency, all participants received an identical set of six images randomly sampled from the AFHQ Wildlife dataset.

In the GANSlider interface, we used the top eight principal components, mapped to sliders in descending order of explained variance. In \sysname, we also used eight dimensions, represented through local PCA scatterplots arranged as pairs of consecutive principal components, again ordered by explained variance.
Like GANSlider, \sysname does not provide real-time numerical feedback on image distance or reconstruction accuracy.

\smallskip\noindent{\textbf{Task Questionnaire:}}
Participants completed a post-task questionnaire to provide subjective feedback. 
The questionnaire includes five NASA-TLX  and four Likert-scale questions assessing usability and effectiveness, and three open-ended questions for qualitative insights.

\subsection{Participants} 
We recruited 15 participants for a user experience study. 
Participants were university students with a variety of backgrounds, including computer science, mathematics, art, graphic design, and human-computer interaction, reflecting a range of perspectives relevant to both technical and creative domains.
The study duration averaged 55 minutes per session, with 30 minutes allocated for training and 25 minutes for trials. 

\subsection{Procedure}
\noindent{\textbf{Study Intro:}}
Participants were first presented with an overview of the study objectives, followed by detailed information regarding data collection, privacy policies, and the participant's right to withdraw at any point without penalty. 
After providing consent, participants were introduced to the two interface variants and briefed on the image reconstruction tasks. 
Before beginning the actual tasks, participants were given a short tutorial to interact with both interfaces and familiarize themselves with the controls. 

\smallskip\noindent{\textbf{Training Phase:}}
Participants were given a live demonstration of both tools, followed by hands-on practice using each tool on a separate dataset, selected from the pretrained StyleGAN2 model on AFHQ Cat or AFHQ Dog. 
This phase ensured that participants understood the functionality and interaction mechanisms of both interfaces prior to beginning the main tasks. 

\smallskip\noindent{\textbf{Image Reconstruction Tasks:}}
Each participant completed twelve image reconstruction tasks: six using GANSlider and six using \sysname. 
To mitigate order effects, both the sequence of tool usage and the order in which the six target images were presented were randomized at the beginning of each trial.

To encourage deliberate exploration and minimize performance variability, no time limits were imposed on individual tasks. Participants were free to proceed at their own pace, with each task concluding only when they actively selected either the ``Skip and Next'' or ``Done and Next'' button. 
Similarly, the study was intentionally designed without an overall time limit, enabling participants to complete all twelve tasks without external pacing pressures. This approach was intended to foster sustained cognitive engagement and reduce confounding effects associated with time-induced stress or fatigue.

\smallskip\noindent{\textbf{Qualitative Feedback:}}
Upon completing all image reconstruction tasks, participants were directed to a survey page where they provided subjective feedback. 
The survey included the five-question NASA-TLX questionnaire to assess workload, four custom Likert-scale statements evaluating usability and satisfaction, and three open-ended questions soliciting general feedback, perceived usefulness, and potential applications of the tools.

\begin{figure}[t]
    \centering
    \includegraphics[width=0.99\linewidth]{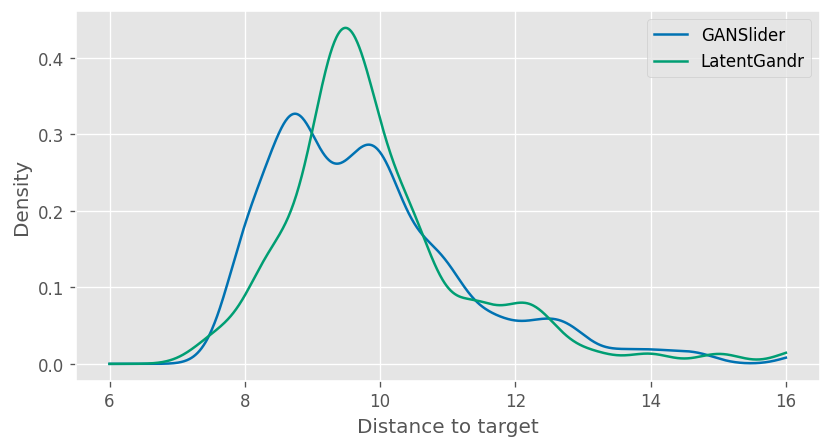}
    \caption{Kernel density estimation of distances to the target images for the two interfaces.}   
    \label{fig:kde-dist}
\end{figure}



\begin{figure}[t]
    \centering
    \includegraphics[width=1\linewidth]{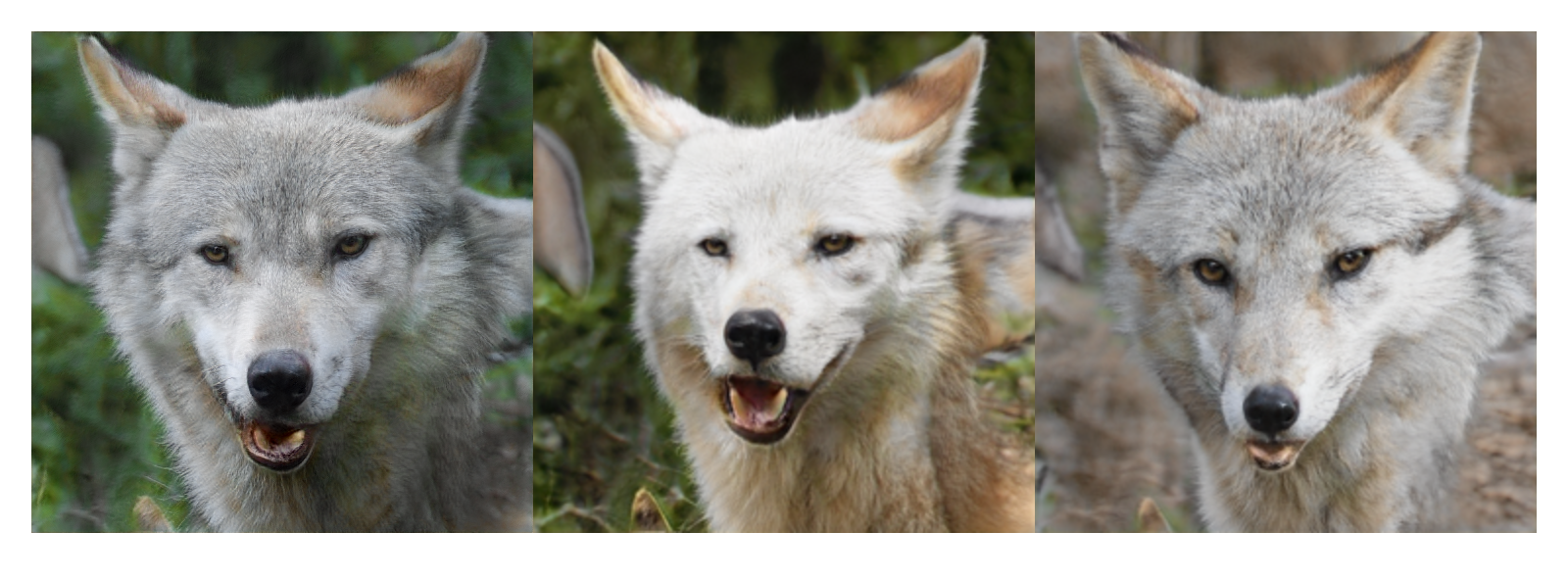}
    \begin{overpic}[width=0.8\linewidth,trim={30mm 10mm 6mm 0mm},clip]{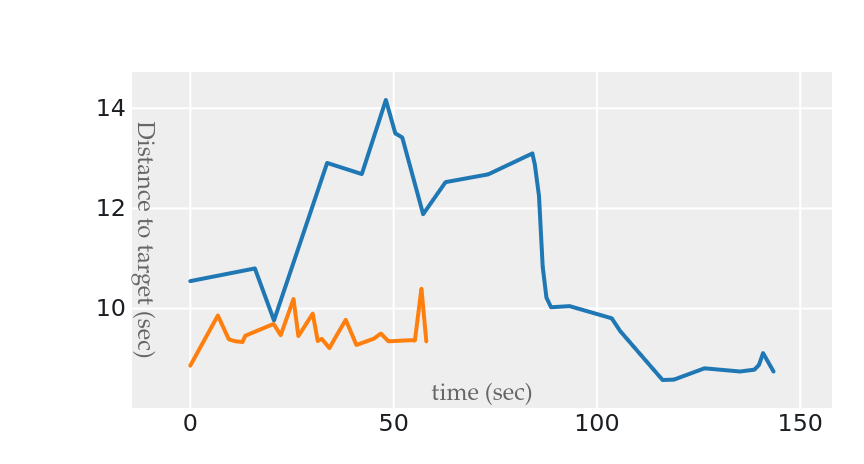}
     \put(46,40){\includegraphics[width=0.4\linewidth,trim={5mm 10mm 195mm 8mm},clip]{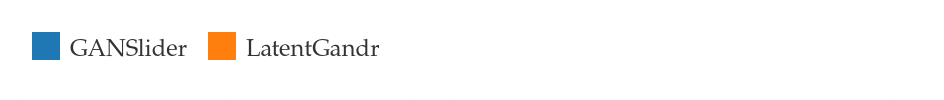}}
    \end{overpic}
    \caption{A comparison of the result. \textbf{Top (left to right)}: target image, image reconstructed by GANSlider and image reconstructed by \sysname
    \textbf{Bottom:} We observed that, although \sysname produced higher measured distances to the target (lower is better), the generated image appeared more visually aligned with the target.}
  
    \label{fig:result-compare}
\end{figure}

\begin{figure}[t]
    \centering
    \includegraphics[width=1\linewidth]{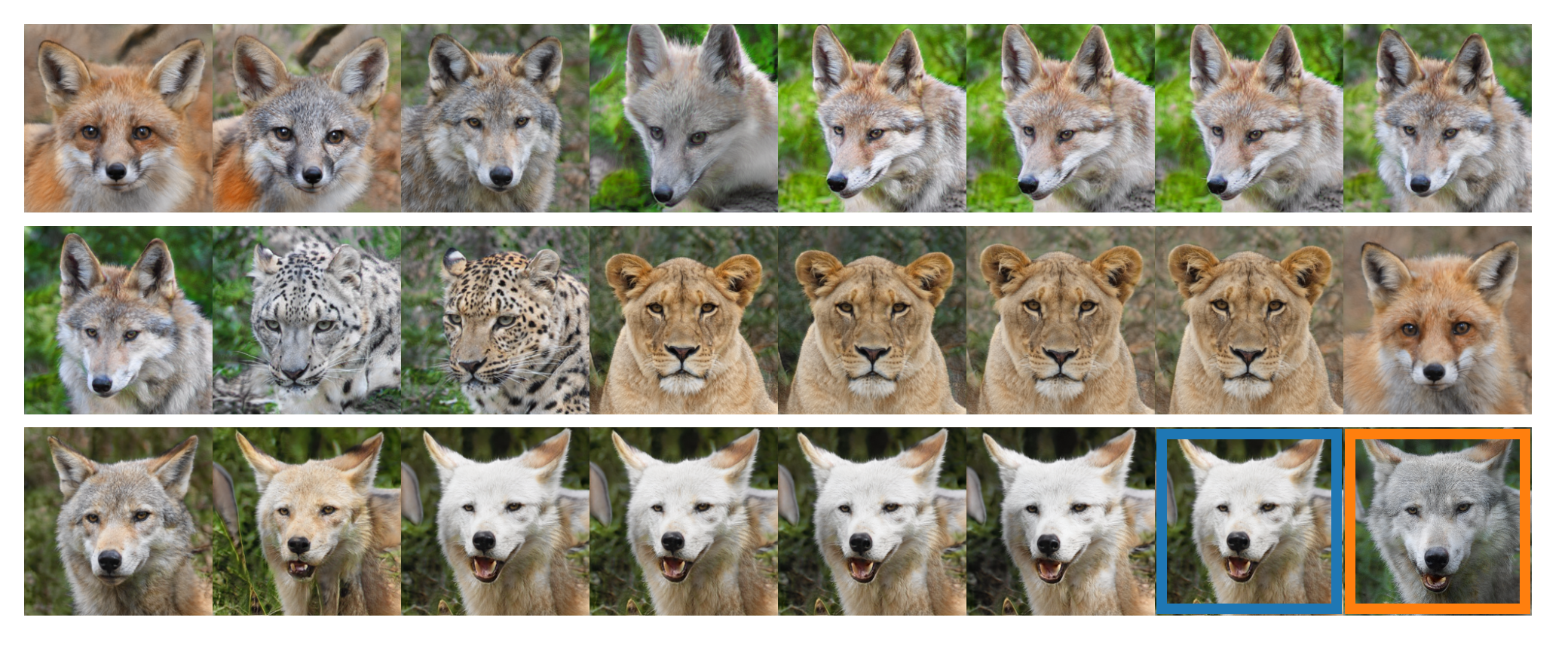}
    \label{fig:seq-gs}
    \includegraphics[width=1\linewidth]{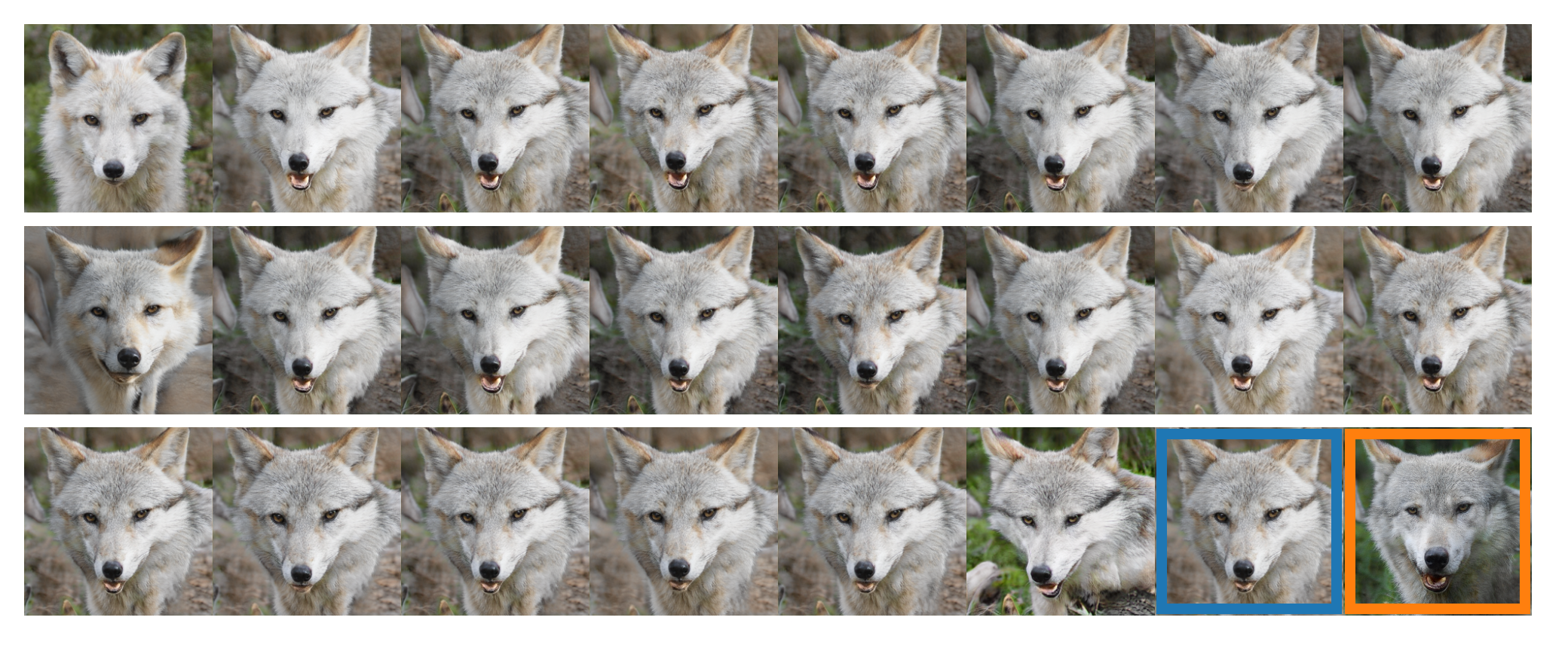}
    \caption{A sample of target-finding sequences in the two interfaces (\textbf{top:} GANSlider, \textbf{bottom:} \sysname ), with the last image representing the target.}
    \label{fig:seq-lg}
    \vspace{-5mm}
\end{figure}

\subsection{Results}
\subsubsection{Task Performance}

Figure~\ref{fig:kde-dist} presents kernel density estimations of the Euclidean distances from reconstructed images to the targets in StyleGAN2's style space. Also see the number of interactions and completion times for the image reconstruction tasks across all participants in supplementary material. 

In terms of reconstruction accuracy, \sysname performed comparably to GANSlider but required more user interactions and longer completion times. Notably, while GANSlider required eight dimensions to capture the variance in the data, \sysname achieved similar performance with fewer dimensions (see Figure~\ref{fig:eval-dist-to-projection}). For consistency, however, both methods were presented using eight dimensions during the study.


Further, we observe that the numerical distance metrics do not capture the full picture. In several instances, despite the higher measured distances, the images generated with \sysname appeared more visually consistent with the target, suggesting potential advantages in maintaining image coherence.
Figure~\ref{fig:result-compare} shows one instance of such cases. 
Although \sysname\ resulted in higher measured distances to the target (with lower values being preferable), the images it generated often appeared more visually aligned with the target.
Examining the reconstruction sequences from both interfaces (Figure~\ref{fig:seq-lg}), we observe that with GANSlider, the user requires several interactions to narrow down to the correct region. In contrast, with \sysname, the user is able to begin fine-tuning immediately after identifying the appropriate neighborhood.

\subsubsection{User Feedback}
To complement our quantitative findings, we administered a post-study questionnaire with three open-ended prompts aimed at probing (1) interpretability, (2) interface improvements, and (3) potential use cases with the following questions:  
\begin{enumerate}[label=Q\arabic*]
    \item Do you have any final remarks on the interpretability of the different slider types?
    \item Is there anything you would add to the interface to improve it?
    \item Can you think of other use cases where either interface would be helpful? 
\end{enumerate}
Qualitative feedback revealed tradeoffs not captured by performance metrics, particularly between expressive control and cognitive load. Participants suggested enhancements such as visual anchors and semantic labels to improve interpretability. The diversity of proposed use cases, ranging from creative design to applied AI, underscores the interface’s broader applicability and informs future design refinements.

\smallskip\noindent\textbf{Interpretability:} 
Participants found both interfaces slightly opaque in terms of which visual features each slider controlled, though differences in workflow influenced perceived interpretability. Participants generally found \sysname more approachable at the outset, offering a clearer visual overview and more confident entry point through the scatter plot. In contrast, GANSlider was viewed as more intuitive during fine-tuning, despite being harder to initialize.

Several participants noted decreasing interpretability when sliders affected multiple features simultaneously, or when semantic meaning (e.g., ``ears'' or ``eyes'') was unclear. Some suggested that labeling sliders or narrowing the image domain (e.g., focusing on one animal type) may improve clarity. A few participants observed that slider interpretability decreased with position (i.e., top to bottom), and that some sliders had overlapping effects, making it harder to isolate specific changes.
Overall, the feedback highlights a tradeoff between global orientation and local precision, and points to a need for clearer feature semantics and disentanglement for more intuitive user control.

\smallskip\noindent{\textbf{Interface improvements:}} 
Participants offered several suggestions to enhance usability, interpretability, and control within the interface. A common theme was the need for better real-time feedback. Users requested features such as a similarity percentage indicator to track progress toward a target image.
Additionally, many highlighted the need for undo functionality, such as a global reset button or single-step undo option. Some also pointed to visual clutter from too many starting images and suggested reducing visual load or filtering by category to streamline initial selection.

\smallskip\noindent\textbf{Use cases:}
Participants identified a range of potential applications for both interfaces. Several noted that the slider-based interface is well-suited for tasks requiring fine-tuning or making incremental adjustments, such as image editing or graphic design. In contrast, the grid or scatter plot interface was seen as effective for establishing an initial configuration or broader exploration across diverse options. Some participants suggested a hybrid workflow, beginning with a scatter plot for broad, initial selection followed by sliders for refinement—one participant observed, ``Sliders [are] nice for fine-tuning while scatter plots would be nice for choosing a starting point.'' Another participant described \sysname as ``start[ing] from a good initial point and then allow[ing] you to [fine-tune] after selecting a solid start.'' These insights reinforce the complementary roles of global and local interactions, and highlight opportunities for hybrid interface design. 


\section{Discussion and Future Work}

The results from our evaluation and user study highlight the strengths and limitations of \sysname in comparison to GANSlider. One of the key advantages of \sysname is its ability to preserve local geometric relationships through localized principal component analysis (Local PCA). This localized approach mitigates the distortions caused by global PCA-based methods, leading to more intuitive and accurate control over the generative process. Participants reported greater confidence in navigating the latent space and found the interactive visualization of Local PCs helpful for understanding how different dimensions influence the generated images.

The reduced likelihood of hallucination in \sysname further enhances its reliability. By constraining exploration to locally linear dimensions, the interface lowers the risk of out-of-sample outputs, a problem more frequently encountered in global PCA-based approaches. 
Additionally, the adaptive neighborhood selection in \sysname ensured that relevant features were prioritized for user interaction, reducing cognitive load and promoting more effective decision-making.

Despite these strengths, there remain opportunities for improvements in \sysname .
While the scatterplot-based visualization offered clear insights into local neighborhoods, some participants expressed a preference for a combination of scatterplots and sliders for more granular control. 
A hybrid interface that integrates global overview visualizations with local refinement options could provide an optimal balance of exploration and precision. 
Participants also recommended enhancing interpretability by labeling sliders or providing additional feedback on how changes in latent dimensions map visually. 

Furthermore, while \sysname's reliance on local neighborhoods improved reconstruction accuracy in theory, it introduced a tradeoff between reconstruction accuracy and the user's cognitive load. 
The increase in the number of local neighborhoods, while ensuring reconstruction accuracy, can become computationally resource-intensive for large-scale datasets or complex generative models.
For users, the search for optimal local neighbors for a target image can become tedious. 
Future work could explore these tradeoffs to mitigate these limitations.

Finally, the generalizability of \sysname was evident from the range of potential use cases identified by the participants. 
Beyond creative design and image generation, the interface showed promise for applications in areas like medical imaging, product design, and simulation-based decision-making. 
This suggests that localized exploration interfaces have broad applicability across domains where interpretability and control are crucial.

Looking ahead, we plan to further refine \sysname by integrating hybrid interaction mechanisms that combine scatterplots and sliders for more flexible control.
Additionally, incorporating real-time feedback on the effects of user adjustments could further enhance the interpretability and usability of the system. 
In the future, we aim to explore the application of \sysname to other generative models, including diffusion models and multi-modal AI systems.

\section{Conclusion}
In this paper, we introduced \sysname, a visual analytics technique designed to enhance the exploration and control of generative AI models through localized principal component analysis. 
By identifying and visualizing local neighborhoods within the latent space, \sysname provides users with an intuitive and effective means of navigating high-dimensional generative models.
Through a series of evaluations, we demonstrated that \sysname outperforms Global PCA-based methods like GANSlider, while preserving local geometric structures. 
Overall, \sysname represents a step toward more accessible and effective human-AI collaboration in creative and analytical tasks, providing users with greater control and deeper insights into the generative process.


\bibliographystyle{abbrv-doi-hyperref}
\newpage
\bibliography{bib}

\end{document}